\documentclass[a4paper,11pt]{article}
\usepackage[dvips]{graphicx}
\usepackage{fancyhdr}
\usepackage{latexsym}
\usepackage{amsfonts}
\usepackage{amssymb}
\usepackage{amsthm}
\usepackage{amsmath}
\usepackage{color}
\usepackage{colordvi}
\usepackage{axodraw}
\usepackage{array}
\usepackage{hyperref}
\usepackage{slashed}
\usepackage[utf8x]{inputenc}
\usepackage{subfigure}
\def\RCT{R\chi T}
\def\t{\tau}

\def\lmas{\ell^+}
\def\lmenos{\ell^-}
\def\pt{p_\tau}
\def\CPT{\chi PT}
\begin{document}
\begin{titlepage}
\begin{flushright}
{UAB-FT-735}
\end{flushright}
\vskip 1.5cm

\begin{center}
{\LARGE \bf The weak radiative pion vertex in $\tau^-\to\pi^-\nu_\tau\ell^+\ell^-$ decays}\\[50pt]
{\sc A. Guevara$^{1}$, G. L\'opez Castro$^{1}$, and P. Roig$^{2}$,}

\vspace{1.4cm} ${}^1$ Departamento de F\'{\i}sica, Centro de Investigaci\'on y de Estudios Avanzados,\\
Apartado Postal 14-740, 07000 M\'exico D.F., M\'exico.\\[15pt]
${}^2$ Grup de F\'{\i}sica Te\`orica, Institut de F\'{\i}sica d'Altes Energies,
Universitat Aut\`onoma de Barcelona, E-08193 Bellaterra, Barcelona, Spain.\\[10pt]
\end{center}

\vfill

\begin{abstract}
We carry out a detailed study of the branching fractions and lepton pair invariant-mass spectrum of $\tau^-\to\pi^-\nu_\tau\ell^+\ell^-$ decays ($\ell=e,\,\mu$). In addition 
to the model-independent ($QED$) contributions, we include the structure-dependent ($SD$) terms, 
which encode information on the hadronization of $QCD$ currents. The form factors describing the $SD$ contributions are evaluated by supplementing Chiral Perturbation Theory 
with the inclusion of the lightest multiplet of spin-one resonances as active degrees of freedom. The Lagrangian couplings have been determined demanding the known 
$QCD$ short-distance behaviour to 
the relevant Green functions and associated form factors in the limit where the number of colours goes to infinity. As a result, we predict 
$BR\left(\tau^-\to\pi^-\nu_\tau e^+e^-\right)=\left(1.7^{+1.1}_{-0.3}\right)\cdot 10^{-5}$ and 
$BR\left(\tau^-\to\pi^-\nu_\tau \mu^+\mu^-\right)\in \left[0.03,1.0\right] \cdot10^{-5}$. According to this, the first decay could be measured in the near future, 
which is not granted for the second one.
\end{abstract}
\vspace*{3.0cm}

PACS~: 13.35.Dx, 12.39.Fe, 12.38.-t
\\
\hspace*{0.45cm} Keywords~: Hadronic tau decays, Chiral Lagrangians, Quantum Chromodynamics. \vfill

\end{titlepage}

\section{Introduction}\label{Intro}

The hadronic final states that can be produced in $\tau$ lepton decays, provide a clean environment to study the dynamics of strong interactions at energies below the $\tau$ 
lepton mass. The leading weak interactions that drives the flavor transitions in these decays are dressed by the strong and electromagnetic interactions to generate a large 
diversity of hadronic and photonic states. The  hadronic vertices can be cleanly extracted and used to test several properties of QCD and electroweak interactions, or to 
extract fundamental parameters of the Standard Model~\cite{Pich:2013kg}. 

  In this paper we study the $\tau^{\pm} \to \pi^{\pm}\nu_{\tau}\ell^+\ell^-$ ($\ell =e$ or $\mu$) decays, which have been considered previously \cite{Dib:2011hc} in 
the context of sterile neutrino exchange overlooking the Standard Model contribution which, to our knowledge, has not been studied before. We will present the results of this 
calculation and analyze the associated phenomenology in this article, ignoring all possible new physics contributions. These decay channels have not been attempted to measure 
so far although, as we will show, they are likely to be detected in the near future facilities. 
The $\tau$ lepton decays under consideration are the crossed channels of the $\pi^{\pm} \to \ell^{\pm} \nu_{\ell} e^+e^-$ decays, which have been studied 
in the past \cite{Bryman:1982et, Kersch:1985rn} and have been already observed \cite{Beringer:1900zz}. Both decays are interesting because they involve the 
$\gamma^*W^{*\mp}\pi^{\pm}$ vertex with the two gauge bosons off their mass-shells. The analogous radiative $\tau^{\pm} \to \pi^{\pm}\nu_{\tau}\gamma$ and 
$\pi^{\pm} \to \ell^{\pm} \nu_\ell\gamma$ decays, which have been widely studied before \cite{Bijnens:1992en, radpidec, radtaudec, Guo:2010dv}, provide information on the same 
vertex in the case of a real photon. The knowledge of the $\gamma W \pi$ vertex in the full kinematical range is of great importance, not only for testing QCD predictions, 
but also because it plays a relevant role in computing the radiative corrections to $\pi \to \ell \nu$, $\tau \to \pi \nu_{\tau}$ decays or in the evaluation of the hadronic 
light-by-light contributions to the muon anomalous magnetic moment \cite{Hlbl}.

These four-body decays of pions and $\tau$ leptons explore different virtualities of the photon and $W$ boson and can provide complementary information on the relevant form 
factors. The low energies involved in pion decays are sensitive to $QCD$ predictions in the chiral and isospin limits, while $\tau$ lepton decays involve energy scales where 
the resonance degrees of freedom become relevant. As is well known, rigorous predictions from $QCD$ for the form factors that describe the $\gamma W\pi$ and $\gamma \gamma \pi$ 
vertices can be obtained only in the chiral and short-distance limits. Therefore the information provided by $\tau$ lepton decays is valuable in order to understand the 
extrapolation between these two limiting cases.

   The vector and axial-vector form factors relevant to our study are calculated in the framework of the Resonance Chiral Theory ($\RCT$) \cite{Ecker:1988te, RChT}. In order 
to fix the free couplings appearing in these calculations we also impose available short-distance constraints in the large $N_C$ limit of $QCD$. As a result, we are able to 
predict the branching ratios and the invariant-mass spectrum of the lepton pair in $\tau^{\pm} \to \pi^{\pm}\nu_{\tau}\ell^+\ell^-$ decays.

   In Sec. \ref{ME and DR} we decompose the matrix element in terms of the model-independent ($QED$) and the $SD$ (vector and axial-vector) contributions, where the latter 
depend on the corresponding hadronic form factors. These are studied in detail in Sec. \ref{SD FFs} and the QCD constraints on their short-distance behaviour 
in the $N_C\to\infty$ limit are discussed in Sec. \ref{Shortdistance}. The related phenomenological analysis is presented in Sec. \ref{Pheno} and we give 
our conclusions in Sec \ref{Concl}. An appendix with the results of the spin-averaged squared matrix element completes our discussion.

\section{Matrix element and decay rate}\label{ME and DR}
\begin{figure}[h!]
\centering
\includegraphics[scale=0.5]{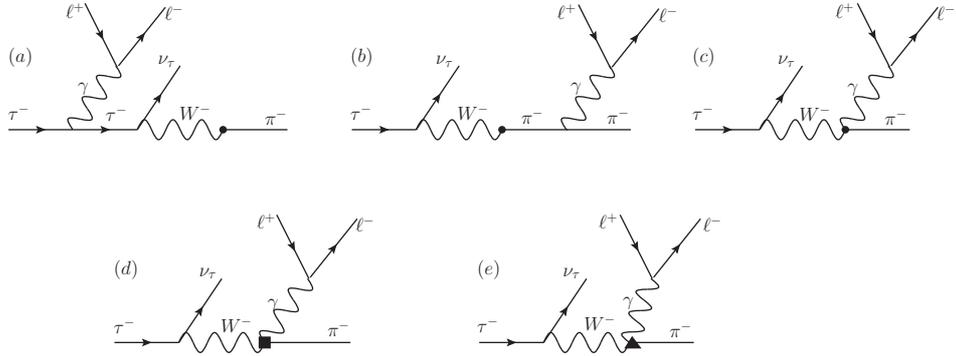}
\caption{Feynman diagrams for the different kinds of contributions to the $\tau^-\to\pi^-\nu_\tau\ell^+\ell^-$ decays, as explained in the main text. The
 dot indicates the hadronization of the $QCD$ currents. The solid square (triangle) represents the $SD$ contribution mediated by the axial-vector (vector) current.}
\label{Fig:1}
\end{figure}

We consider the process $\tau^-(p_\tau)\to\pi^-(p)\nu_\tau(q)\ell^+(p_+)\ell^-(p_-)$. This decay is generated by demanding that the photon in the $\tau^-(p_\tau)\to\pi^-(p)\nu_\tau(q)\gamma(k)$ 
decays becomes virtual and then converts into a lepton pair (lepton pair production mediated by the $Z$ boson is negligible); at the amplitude level, it suffices to 
change the photon polarization $\epsilon^{\mu}$ in the radiative decay by $e\bar{u}(p_-)\gamma_{\mu}v(p_+)/k^2$, with $k=p_++p_-$ the photon momentum and $e$ the positron 
charge. Therefore, one can relate the description of the structure dependent contributions in the former to that in the latter \cite{Guo:2010dv}. In analogy with the 
radiative pion and one-meson tau decays, the matrix element can be written as the sum of four contributions:
\begin{equation}\label{matrix element decomposition}
   \mathcal{M} \left[\tau^-(p_\t) \to \pi^-(p) \nu_\tau(q) \lmas(p_+)\lmenos(p_-)\right]
   = \mathcal{M}_{IB_\tau} + \mathcal{M}_{IB_\pi} + \mathcal{M}_{V} + \mathcal{M}_{A}\,.
\end{equation}
The relevant diagrams are depicted in Fig.\ref{Fig:1}. The notation introduced for the amplitudes describes the four kinds of 
contributions: $\mathcal{M}_{IB_{\tau}}$ is the bremsstrahlung off the tau lepton, (figure \ref{Fig:1}(a)); $\mathcal{M}_{IB_{\pi}}$ is the sum of the bremsstrahlung off the 
$\pi$ meson (figure \ref{Fig:1}(b)), and the diagram with the local $W^*\gamma^*\pi$ vertex (figure \ref{Fig:1}(c)); $\mathcal{M}_{V}$ is the structure dependent vector 
contribution (figure \ref{Fig:1}(d)) and $\mathcal{M}_{A}$ the structure dependent axial-vector contribution (figure \ref{Fig:1}(e)). Our imprecise knowledge of the exact 
mechanism of hadronization in the last two terms is parametrized in terms of hadronic form factors, which are functions of $p\cdot k$ and $k^2$
. 

The decay amplitude is composed of the following set of gauge-invariant contributions ($G_F$ is the Fermi constant, $V_{ud}=0.9742$ the $ud$ quark mixing angle, 
$F_\pi=92.2$ MeV \cite{Beringer:1900zz} and we have defined $ \mathcal{M}_{IB}=\mathcal{M}_{IB_{\tau}}+\mathcal{M}_{IB_{\pi}}$),
\begin{eqnarray}\label{explicit expressions matrix element}
 \mathcal{M}_{IB} & = & -i G_F V_{ud}\frac{e^2}{k^2}F_\pi M_\tau \bar{u}(p_-)\gamma_\mu v(p_+)\bar{u}(q)(1+\gamma_5)\left[\frac{2p^\mu}{2p\cdot k+k^2} + 
\frac{2\pt^\mu-\slashed{k}\gamma^\mu}{-2\pt\cdot k+k^2}\right]u(\pt)\,,\nonumber\\
 \mathcal{M}_{V} & = & - G_F V_{ud} \frac{e^2}{k^2} \bar{u}(p_-)\gamma^\nu v(p_+) F_V(p\cdot k,k^2)\epsilon_{\mu\nu\rho\sigma}k^\rho p^\sigma \bar{u}(q)\gamma^\mu(1-\gamma_5)u(\pt)\,,\\
 \mathcal{M}_{A} & = & i G_F V_{ud} \frac{2e^2}{k^2}\bar{u}(p_-)\gamma_\nu v(p_+) \left\lbrace F_A(p\cdot k,k^2)\left[(k^2+p\cdot k)g^{\mu\nu}-k^\mu p^\nu\right] - \frac{1}{2} A_2(k^2) k^2g^{\mu\nu} \right. \nonumber \\ && \ \ \ \ \left. + \frac{1}{2} A_4(k^2) k^2(p+k)^\mu p^\nu\right\rbrace\bar{u}(q)\gamma_\mu(1-\gamma_5)u(\pt)\nonumber\, .
\end{eqnarray}
The structure-dependent contributions are described in terms of one vector and three axial-vector Lorentz invariant form factors. These form factors will be discussed in detail 
later in the article and, in particular, the dependence on $k^2$ of $F_A(p\cdot k,k^2)$ and $F_V(p\cdot k,k^2)$ will be given in section \ref{SD FFs}. It can be easily 
checked that the decay amplitudes corresponding to the radiative $\tau^- \to \pi^-\nu_\tau\gamma$ decays can be obtained from  Eq. (\ref{explicit expressions matrix element}) 
by replacing $e\bar{u}(p_-)\gamma^{\mu}v(p_+) \rightarrow \epsilon^{\mu}/k^2$, where $\epsilon^{\mu}$ is the polarization four-vector of the real photon, and then by setting 
$k^2=0$. In this case, the decay amplitude depends only upon two form factors, $F_A(p\cdot k,k^2=0)$ and $F_V(p\cdot k,k^2=0)$, whose expressions can be read from 
Ref.~\cite{Guo:2010dv}. The additional axial-vector form factors $A_2(k^2)$, and $A_4(k^2)$ can be found in Ref.~\cite{Bijnens:1992en}.

Eq.(\ref{explicit expressions matrix element}) can be checked from the corresponding expressions for $K^+ \to \mu^+\nu_\mu\ell^+\ell^-$ in 
eq.(4.9) in Ref.~\cite{Bijnens:1992en} by using crossing symmetry and the conservation of the 
electromagnetic current. As noted in this reference, the parametrization of the axial-vector form factor used by the Particle Data Group \cite{Beringer:1900zz} for the 
analogous $\pi^+ \to \mu^+\nu_{\mu} e^+e^-$ decays, neglects the $A_4(k^2)$ form factor \footnote{The other form factors are related via 
$-\sqrt{2}m_\pi\left[F_A(p\cdot k,k^2),A_2(k^2),F_V(k^2)\right]=\left[F_A,R,F_V\right]$ to the ones used in Ref.~\cite{Beringer:1900zz}.}. Given the different kinematics of 
our problem we will keep it in the following. 
As we will see later, at next-to-leading order in $\chi PT$, $A_2(k^2)$ and $A_4(k^2)$ can be expressed in terms of only one form factor (this is no longer true 
at the next order \cite{Bijnens:1992en}, whose contributions we neglect). If we define this form factor as $B(k^2)\equiv-\frac{1}{2}A_2(k^2)$, then 
$\frac{1}{2}A_4(k^2)=-B(k^2)/(k^2+2p\cdot k)$ and the axial-vector SD amplitude is simplified to
\begin{eqnarray}\label{A matrix element}
  \mathcal{M}_{A} & = & i G_F V_{ud} \frac{2e^2}{k^2}\bar{u}(p_-)\gamma_\nu v(p_+) \left\lbrace F_A(p\cdot k,k^2)\left[(k^2+p\cdot k)g^{\mu\nu}-k^\mu p^\nu\right]\right.\nonumber\\
& & \left. +B(k^2) k^2 \left[g^{\mu\nu}-\frac{(p+k)^\mu p^\nu}{k^2+2p\cdot k}\right]\right\rbrace\bar{u}(q)\gamma_\mu(1-\gamma_5)u(\pt)\,.
\end{eqnarray}

The results of summing the different contributions to the squared matrix element over polarizations are collected in the appendix.

The $IB$ contributions are model-independent in the sense that they are determined in terms of the parameters of the well known non-radiative $\tau^- \to \pi^- \nu_\tau$ decays 
and using $QED$. They provide the dominant contribution to the decay rate in the case of a real photon emission  \cite{Guo:2010dv} owing to the well known infrarred divergent behavior. For the decay 
under consideration we can expect that this behaviour is softened since $k^2 \geq 4m_{\ell}^2$.  The $SD$ (or model-dependent) contributions require the modelling of the 
$\gamma^* W^*\pi$ vertex for photon and $W$ boson virtualities of the order of $1$ GeV. Those terms can be split into a vector $V$ and an axial-vector $A$ contributions 
according to Eq. (\ref{explicit expressions matrix element}) and must include the resonance degrees of freedom that are relevant at such energies (see Sec. \ref{SD FFs}).

Therefore, the decay rate can be conveniently separated into six terms which correspond to three moduli squared ($IB,\ VV,\ AA$) and three interference terms ($IB-V,\ IB-A,\ 
V-A$). Thus, we can write the decay rate as follows:
\begin{eqnarray} \label{parts Gamma radiative decay tau one pG}
    \Gamma_{\rm total} =  \Gamma_{IB} + \Gamma_{VV} + \Gamma_{AA}+  \Gamma_{IB-V} + \Gamma_{IB-A}+\Gamma_{V-A}\ .
\end{eqnarray}
In terms of the five independent kinematical variables needed to describe a four-body decay, the differential decay rate is given by
\begin{equation}\label{differential decay rate}
 {\rm d}\Gamma\left(\tau^-\to\nu_\tau\pi^-\ell^+\ell^-\right) = \frac{X \beta_{12}\beta_{34}}{4(4\pi)^6M_\tau^3}\overline{|\mathcal{M}|^2} {\rm d}s_{34} {\rm d}s_{12} 
{\rm d}({\rm cos}\theta_1) {\rm d}({\rm cos}\theta_3)  {\rm d}\phi_3\,,
\end{equation}
where $\overline{|\mathcal{M}|^2}$ is the spin-averaged unpolarized decay probability,
\begin{equation}\label{definitions differential decay rate}
 X = \frac{\lambda^{1/2}(M_\tau^2,s_{12},s_{34})}{2}\,,\quad \beta_{ij}\,=\,\frac{\lambda^{1/2}(s_{ij},m_i^2,m_j^2)}{s_{ij}}\,,
\end{equation}
and $\lambda(a,b,c)=a^2+b^2+c^2-2ab-2ac-2bc$.
 
The five independent kinematical variables in eq. (\ref{differential decay rate}) were chosen as 
$\left\lbrace s_{12},\,s_{34},\,\theta_1,\,\theta_3,\,\phi_ 3\right\rbrace$, where $s_{12}:=(p_1+p_2)^2$ and $s_{34}:=(p_3+p_4)^2$; the momenta were relabelled 
\footnote{We decided to write eqs.(\ref{differential decay rate}) and (\ref{definitions differential decay rate}) in terms of the second set of momenta in 
eq.(\ref{relabelling}) for its general usefulness in four-body decays. See Ref.\cite{AlainGabriel} for details. On the contrary, we prefer to present the rest of 
eqs.(\ref{matrix element decomposition} to (\ref{averaged VA}) in terms of the first set of momenta in eq.(\ref{relabelling}) for an easier interpretation.} as 
\begin{equation}\label{relabelling}
 \left\lbrace p_\tau,\, q,\, p,\, p_+,\, p_-\right\rbrace \to \left\lbrace p,\, p_1,\, p_2,\, p_3,\, p_4\right\rbrace.
\end{equation}
The definition of the angles is the standard one. Finally, the integration limits are
\begin{eqnarray}\label{integration limits}
& & s_{34}^{min}=(m_3+m_4)^2\,,\;s_{34}^{max}=(M-m_1-m_2)^2\,,\quad \theta_{1,3}\in [0,\pi]\,,\quad \phi_3\in[0,2\pi]\,,\nonumber\\
& & s_{12}^{min}=(m_1+m_2)^2\,,\;s_{12}^{max}=\left(M-\sqrt{s_{34}}\right)^2\,.
\end{eqnarray}
In this way, the outermost integration corresponds to the square of the invariant mass $s_{34}$ of the lepton-antilepton pair, assuming it can be the easiest spectrum to be measured in the 
considered decays.

\section{Structure-dependent form factors}\label{SD FFs}
Although the hadronic form factors cannot be computed from the underlying theory, the symmetries of QCD are nonetheless the guiding principle to write the effective Lagrangian 
that will be used. At very low energies, the strong interaction Lagrangian exhibits a chiral $SU(n_f)\otimes SU(n_f)$ symmetry in the approximate limit of ($n_f$) massless light quarks. 
This symmetry allows to develop $\CPT$ \cite{CPT} as an expansion in powers of momenta and masses of the lightest mesons (that acquire mass through explicit chiral symmetry breaking), 
over a typical hadronic scale which can be identified with the lightest resonances or the chiral symmetry breaking scale. Since the energies probed in hadronic tau decays are 
larger than these hadronic scales, the $\CPT$ expansion parameter does no longer converge at high invariant masses. In paralel new degrees of freedom, the lightest resonances, 
become excited and they should be introduced as dynamical fields in the action. This is done in $\RCT$ \cite{Ecker:1988te} working in the convenient antisymmetric tensor 
formalism which warrantees that the contact interactions of next-to-leading order ($NLO$) $\CPT$ are already included in the $\RCT$ Lagrangian, as can be seen by integrating 
the resonances out. Now the expansion parameter is $1/N_C$ ($N_C$ being the number of colours of the gauge group) \cite{LargeN} and the theory at leading order has a spectrum 
of infinitely many stable states with only tree level interactions. In our case, we will see that the kinematics of the problem damps very strongly the observables above $1$ 
GeV, which justifies considering only the exchange of the lightest vector and axial-vector resonance multiplets \footnote{Given the (axial-)vector character of the Standard 
Model couplings of the hadronic matrix elements in $\tau$ decays, form factors for these processes are ruled by vector and axial-vector resonances.}. We will introduce the most 
important $NLO$ correction in the $1/N_C$ counting given by the meson widths as they are needed to achieve a sensible description of the propagating resonances.

The relevant effective Lagrangian reads~:
\begin{eqnarray}
\label{eq:ret} {\cal L}_{\rm R\chi T}   & \doteq   & {\cal L}_{WZW} \,+ \,
{\cal L}_{\rm kin}^{\rm V}\, + \, \frac{F_\pi^2}{4}\langle u_{\mu} u^{\mu} + \chi _+
\rangle \, + \, \frac{F_V}{2\sqrt{2}} \langle V_{\mu\nu} f_+^{\mu\nu}\rangle
\, \nonumber \\
& &  \hspace{-1.9cm} + \ i \,\frac{G_V}{\sqrt{2}} \langle V_{\mu\nu} u^\mu
u^\nu\rangle \, +\, \sum_{i=1}^{7}  \, \frac{c_i}{M_V}  \, {\cal O}^i_{\rm
VJP} \,+\, \sum_{i=1}^{4}  \, d_i  \, {\cal O}^i_{\rm VVP} \,+\,
\sum_{i=1}^{5}  \, \lambda_i  \, {\cal O}^i_{\rm VAP} \ ,
\label{lagrangian}
\end{eqnarray}
where all coupling constants are real and $M_V$ is the mass of the lightest vector meson resonance nonet \cite{Cirigliano:2003yq}.
We follow here the notation in Refs.~\cite{Ecker:1988te,RuizFemenia:2003hm,GomezDumm:2003ku}, where the explicit form of these operators can be found.

\begin{figure}[ht]
\begin{center}
\includegraphics[scale=0.6]{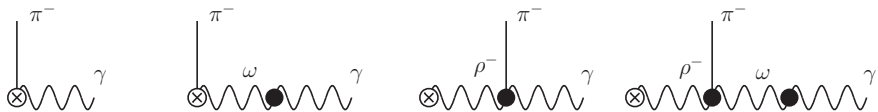}
\caption{Vector current contributions to the $W^{-*}\rightarrow \pi^- \gamma^*$ vertex. \label{fig.v}}
\end{center}
\end{figure}

\begin{figure}[ht]
\begin{center}
\includegraphics[scale=0.6]{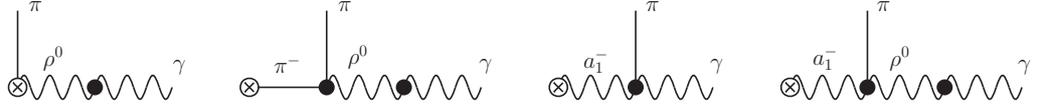}
\caption{Axial-vector current contributions to the $W^{-*}\rightarrow \pi^- \gamma^*$ vertex. \label{fig.a}}
\end{center}
\end{figure}

The structure-dependent form factors in $\tau^-\to\nu_\tau\pi^-\ell^+\ell^-$ decays that appear in Eq. (\ref{explicit expressions matrix element}), can be obtained from the 
same Feynman diagrams considered in Ref.~\cite{Guo:2010dv} for the $\tau^-\to\nu_\tau\pi^-\gamma$ decays. This is achieved by replacing the real photon by a virtual one, 
which then converts into the lepton-antilepton pair. These diagrams are given in Figs.\ref{fig.v} and \ref{fig.a} for the vector and axial-vector current contributions, 
respectively.\\
\\
\\
Since both the $W$ gauge boson and the photon are virtual in the present case, the form factors defining the $\gamma^*W^* \pi$ vertex will depend upon two invariant variables 
which we choose as  $t:=(p+k)^2=k^2+2p\cdot k+m_\pi^2$ and $k^2$. The other important difference is that the second diagram of Fig.\ref{fig.a} -which was zero for real 
photons~\cite{Guo:2010dv}- will now contribute, giving rise to the additional form factor $B\left(k^2\right)$. This term can be related to the isovector component of the 
electromagnetic $\pi^+\pi^-$ form factor \cite{Bijnens:1992en} and it accounts for the off-shellness of the photon that is not contained in the pure scalar $QED$ contribution.

In the framework of the $\RCT$ the vector form factor $F_V\left(t, k^2\right)$, defined in eq.(\ref{explicit expressions matrix element}), adopts the following expression:
\begin{eqnarray}\label{F_V}
 F_V(t,k^2) &=& -\frac{N_C}{24\pi^2 F_\pi}+ \frac{2\sqrt2 F_V}{3 F_\pi M_V
}\bigg[ (c_2-c_1-c_5) t +
(c_5-c_1-c_2-8c_3) m_\pi^2 + 2 (c_6-c_5) k^2\bigg]\times\nonumber \\
& &  \left[ \frac{\mathrm{cos}^2\theta}{M_\phi^2-k^2-iM_\phi\Gamma_\phi}\left(1-\sqrt{2} \mathrm{tg}\theta \right)
+ \frac{\mathrm{sin}^2\theta}{M_\omega^2-k^2-iM_\omega\Gamma_\omega}\left(1+\sqrt{2} \mathrm{cotg}\theta \right)\right]
\nonumber \\
& & + \frac{2\sqrt2 F_V}{3 F_\pi M_V }\, D_\rho(t)\,  \bigg[ ( c_1-c_2-c_5+2c_6) t +
(c_5-c_1-c_2-8c_3) m_\pi^2 + (c_2-c_1-c_5)k^2\bigg] \nonumber \\
& & + \frac{4 F_V^2}{3 F_\pi }\, D_\rho(t)\,  \bigg[ d_3 (t+4k^2) +
(d_1+8d_2-d_3) m_\pi^2 \bigg]\times\nonumber \\
& & \left[ \frac{\mathrm{cos}^2\theta}{M_\phi^2-k^2-iM_\phi\Gamma_\phi}\left(1-\sqrt{2} \mathrm{tg}\theta \right)
+ \frac{\mathrm{sin}^2\theta}{M_\omega^2-k^2-iM_\omega\Gamma_\omega}\left(1+\sqrt{2} \mathrm{cotg}\theta \right)\right]\,, 
\end{eqnarray}
where
\begin{equation}
D_\rho(t) = \frac{1}{M_\rho^2 - t - i M_\rho \Gamma_\rho(t)}\,\,,
\end{equation}
and $\Gamma_\rho(t)$ stands for the decay width of the $\rho(770)$ resonance included following the definition given in Ref. \cite{GomezDumm:2000fz}:
\begin{equation}\label{rhowidth}
 \Gamma_\rho(s)=\frac{sM_\rho}{96\pi F_\pi^2}\left[\sigma_\pi^3(s)\theta\left(s-4m_\pi^2\right)+\frac{1}{2}\sigma_K^3(s)\theta\left(s-4m_K^2\right)\right]\,,
\end{equation}
with $\sigma_P(s)=\sqrt{1-\frac{4m_P^2}{s}}$.

For the purposes of numerical evaluation, we will assume the ideal mixing for the $\omega-\phi$ system of vector resonances, namely:
\begin{eqnarray}
\omega_1 = \mathrm{cos}\theta \;\omega - \mathrm{sin}\theta\;\phi \; \sim \sqrt{\frac{2}{3}} \omega - \sqrt{\frac{1}{3}} \phi \,, \nonumber \\
\omega_8 = \mathrm{sin}\theta \;\omega + \mathrm{cos}\theta\;\phi \; \sim \sqrt{\frac{2}{3}} \phi + \sqrt{\frac{1}{3}} \omega \, .
\end{eqnarray}
In this limit, the contribution of the $\phi$ meson to eq.(\ref{F_V}) vanishes; in addition we will neglect any energy-dependence in their off-shell widths given that they 
are rather narrow resonances.

Similarly, the axial-vector form-factor $F_A(t, k^2)$ is given by
\begin{eqnarray} \label{F_A}
F_A(t, k^2) &=& \frac{F_V^2}{F_\pi}\left(1-\frac{2G_V}{F_V}\right)\,D_\rho(k^2) - \frac{F_A^2}{F_\pi} D_{\mathrm{a}_1}(t)
+ \frac{F_A F_V}{\sqrt{2} F_\pi}\,D_\rho(k^2)\, D_{\mathrm{a}_1}(t)\,  \bigg( - \lambda'' t +
\lambda_0 m_\pi^2 \bigg)\,,\nonumber\\
\end{eqnarray}
where we have used the notation 
\begin{eqnarray}
\sqrt{2}\lambda_0 &=&-4\lambda_1- \lambda_2-\frac{\lambda_4}{2}-\lambda_5\,, \nonumber \\
\sqrt{2} \lambda''  &=& \lambda_2-\frac{\lambda_4}{2}-\lambda_5\,,
\end{eqnarray}
for the relevant combinations of the couplings in $\mathcal{L}_2^{VAP}$, eq(\ref{lagrangian}).

The energy-dependent a$_1(1260)$ resonance width entering $D_{\mathrm{a}_1}(t)$ was studied within this framework in Ref.~\cite{Dumm:2009va} where the dominant 
$\pi\pi\pi$ and $KK\pi$ absorptive cuts where obtained in terms of the corresponding three-meson form factors \cite{Dumm:2009va, Dumm:2009kj}. Here we have used the updated 
fit results of Ref.~\cite{TAUOLA2} which were obtained using the complete multi-dimensional distributions measured by BaBar \cite{Nugent:2013ij}.

Finally, the additional axial-vector form factor $B(k^2)$, is
\begin{equation} \label{B}
 B(k^2) = F_\pi \frac{F_V^{\pi^+\pi^-}|_\rho(k^2)-1}{k^2}\,,
\end{equation}
where $F_V^{\pi^+\pi^-}|_\rho$ corresponds to the $I=1$ part of the $\pi^+\pi^-$ vector form factor. Based on the effective field theory description of Ref.\cite{Guerrero:1997ku} 
including only the $\rho(770)$ contribution and reproducing the $\chi PT$ results \cite{Gasser:1990bv, Bijnens:1998fm, Bijnens:2002hp}, several phenomenological approaches 
including the effect of higher excitations have been developed \cite{SanzCillero:2002bs, Roig:2011iv}. This form factor has also been addressed 
within dispersive representations exploiting analyticity and unitarity constraints \cite{Pich:2001pj, De Troconiz:2001wt, Ananthanarayan:2011xt, Hanhart:2012wi}. Here we will 
follow the approach of Ref.~\cite{Dumm:2013zh} and will use a dispersive representation of the form factor at low energies matched to a phenomenological description at 
intermediate energies including the excited resonances contribution. A three-times subtracted dispersion relation will be used
\begin{equation}\label{FV_3_subtractions}
 F_V^\pi(s) \,=\,\exp \Biggl[ \alpha_1\, s\,+\,\frac{\alpha_2}{2}\,
s^2\,+\,\frac{s^3}{\pi}\! \int^\infty_{s_{\rm thr}}\!\!ds'\,
\frac{\delta_1^1(s')} {(s')^3(s'-s-i\epsilon)}\Biggr] \, ,
\end{equation}
where \cite{Boito:2008fq}
\begin{equation}
\label{delta}
 \tan \delta_1^1(s) = \frac{\Im m F_V^{\pi(0)}(s)}{\Re e
F_V^{\pi(0)}(s)} \ ,
\end{equation}
with 
\begin{eqnarray} \label{SU2formula}
\hspace{-.5cm} F_V^{\pi\,(0)}(s) & = & \frac{M_\rho^2}{M_\rho^2
\left[1+\frac{s}{96\pi^2 F_\pi^2}\left(A_\pi(s)+
\frac12 A_K(s)\right)\right]-s}\nonumber \\
& = & \frac{M_\rho^2}{M_\rho^2 \left[1+\frac{s}{96\pi^2 F_\pi^2}\Re e
\left(A_\pi(s) + \frac12 A_K(s)\right)\right]-s-i M_\rho
\Gamma_\rho(s)}\ .
\end{eqnarray}
The loop function is ($\mu$ can be taken as $M_\rho$)
\begin{equation}\label{loopfun_2pi}
A_{P}(k^2) \, = \ln{\left( \frac{m^2_P}{\mu^2}\right)} + {8 \frac{m^2_P}{k^2}} -
\frac{5}{3}  + \sigma_P^3(k^2) \,\ln{\left(\frac{\sigma_P(k^2)+1}{\sigma_P(k^2)-1}\right)}\,,
\end{equation}
and the phase--space factor $\sigma_P(k^2)$ was defined after Eq. (\ref{rhowidth}).

The parameters $\alpha_1,\,\alpha_2$ and the $\rho(770)$ resonance parameters entering $B(k^2)$ will be extracted \cite{Dumm:2013zh} from fits to BaBar 
$\sigma(e^+e^-\to\pi^+\pi^-)$ data \cite{Aubert:2009ad} excluding the $\omega(782)$ contribution. We have used the preliminary values 
$\alpha_1=1.87,\,\alpha_2=4.26$ in the numerics.

\section{Short-distance constraints}\label{Shortdistance}
The form factors derived in the previous Section satisfy the constraints imposed by chiral symmetry. Some of the remaining free parameters can be fixed by requiring that they 
satisfy the short-distance $QCD$ behavior. The study of two-point spin-one Green functions within perturbative QCD \cite{Floratos:1978jb} showed that both of them go to a 
constant value at infinite transfer of momenta. Assuming local duality, the imaginary part of the quark loop can be understood as the sum of infinite positive contributions 
of intermediate hadron states. If these must add up to a constant it should be expected that each of the contributions vanishes in that limit. This vanishing should be 
accomplished asymptotically and, consequently, it is expected that all resonance excitations up to the QCD continuum contribute to the meson form factors in this limit. This 
conclusion is also derived from the large-$N_C$ limit of QCD, where these requirements find their most natural application.

On the contrary, phenomenology suggests that the effect of excited resonances on the short-distance relations is pretty small. To give just two examples, if the effects of 
the $\rho(1450)$ resonance are ignored in the pion vector form factor \cite{Guerrero:1997ku}, the generic asymptotic constraint ($i$ corresponds to the index of the multiplet)
\begin{equation}
 \sum_i F_V^i G_V^i = F_\pi^2\,,
\end{equation}
that is obtained in the $N_C\to\infty$ limit reduces to $F_VG_V=F_\pi^2$. Upon integration of the resonances, this produces the prediction of the $\chi PT$ low-energy coupling 
\begin{equation}
L_9=\frac{F_VG_V}{2M_\rho^2}=\frac{F_\pi^2}{2M_\rho^2}=7.2\cdot10^{-3}\,,
\end{equation}
in remarkably good agreement with the phenomenologically extracted value, which shows that the corrections to obtaining the high-energy constraint considering only the lightest 
multiplet are smaller than $5\%$ in this case.

Our second example concerns the study of the $\tau^-\to (K\pi)^-\nu_\tau$ decays. In Ref.~\cite{Jamin:2006tk} the effect of the $K^\star(1410)$ resonance was included through
\begin{equation}
 \gamma = -\frac{F_V'G_V'}{F^2}=\frac{F_VG_V}{F^2}-1\,.
\end{equation}
While $\gamma=0$ if the second multiplet is neglected, in the subsequent analyses \cite{Jamin:2008qg, Boito:2008fq, Boito:2010me} it was found $\gamma=-0.05\pm0.02$, which 
supports the idea that the modifications introduced by the second multiplet to the short distance constraints are at the $5\%$ level \footnote{This conclusion is supported 
by the analysis of the $\tau^-\to K^-\eta\nu_\tau$ decays \cite{Escribano:2013bca}.}.

This number should be, however, enlarged for estimating the error associated to the neglect of the heavier multiplets on the high-energy constraints in our problem. The 
previous examples were given for two-meson form factors and we are dealing with the form factors corresponding to an (axial-vector) current coupled to a pseudoscalar 
and a photon (giving the lepton-antilepton pair), which has a much richer dynamics. Our estimate on the error is discussed at the end of this section. 

The vanishing of the vector form factor in eq.(\ref{F_V}) for $t\to\infty$ and $k^2\to\infty$ yields
\begin{eqnarray}
 c_1-c_2+c_5&=&0\,, \nonumber \\
 2(c_6-c_5)&=&\frac{-N_C M_V}{32\sqrt{2}\pi^2F_V}\,,
\end{eqnarray}
in agreement with the results of Ref.~\cite{RuizFemenia:2003hm} for the $VVP$ Green's function. No restrictions are found on the other couplings entering eq.(\ref{F_V}). The 
high-energy conditions found in Ref.~\cite{RuizFemenia:2003hm} for them are
\begin{eqnarray}
 -c_1-c_2-8c_3+c_5&=&0\,,\nonumber \\  d_1+8d_2-d_3&=&\frac{F_\pi^2}{8F_V^2}\,, \\ 
 d_3&=& \frac{-N_C}{64\pi^2}\frac{M_V^2}{F_V^2}+\frac{F_\pi^2}{8F_V^2}\, . \nonumber 
\end{eqnarray}
No short-distance requirements are obtained for the axial-vector form factor in eq.(\ref{F_A}), which already vanishes in the limit of $k^2$ and $t$ simultaneously large. The 
corresponding couplings are constrained by the high-energy conditions on the two-point Green functions of vector and axial-vector currents \cite{Ecker:1988te}
\begin{equation}
 F_V G_V = F_\pi^2\,,\quad 2 F_V G_V - F_V^2=0\,,
\end{equation}
and by the short-distance constraints applying in the $VAP$ Green's function \cite{Cirigliano:2004ue} and three-meson hadronic form factors \cite{Dumm:2009va, Dumm:2009kj}:
\begin{equation}
 \lambda'=\frac{F_\pi^2}{2\sqrt{2}F_AG_V}\,,\quad  \lambda''=\frac{2G_V-F_V}{2\sqrt{2}F_A}\,,\quad \lambda_0=\frac{\lambda'+\lambda''}{4}\,.
\end{equation}
If the Weinberg sum rules \cite{Weinberg:1967kj} ($F_V^2-F_A^2=F_\pi^2$, $F_V^2 M_V^2=F_A^2M_A^2$) are imposed, all couplings are predicted in terms of $F_\pi$ and $M_V$:

\begin{eqnarray}\label{Couplings}
c_1-c_2+c_5&=&0\,,\nonumber \\ 2(c_6-c_5)&=&\frac{-N_C M_V}{64\pi^2F_\pi}\,,\nonumber \\ c_1-c_2-8c_3+c_5&=&0\,, \nonumber \\ d_1+8d_2-d_3&=& \frac{1}{16}\,, \\
 d_3&=&\frac{-N_C M_V^2}{128\pi^2F_\pi^2}+\frac{1}{16}\,,\nonumber \\ G_V&=& \frac{F_\pi}{\sqrt{2}}\,,\nonumber \\ F_V&=&\sqrt{2}F_\pi\,,\nonumber \\ F_A&=& F_\pi\,,\nonumber \\ \lambda'&=& \frac{1}{2}\,,\nonumber \\ \lambda''&=&0 \,,\nonumber \\ \lambda_0&=&\frac{1}{8}\,. \nonumber
\end{eqnarray} 
In numerical evaluations we will take $M_V=775$ MeV.

 In order to estimate the error of our predictions we may be conservative and consider uncorrelated variations of the above relations (\ref{Couplings}) of around $1/3$. 
Comparison to hadronic tau decay data suggests, however, that the typical error of our approach is smaller \cite{Roig:2012zj}, $\lesssim20\%$, and we will take this figure 
for estimating the error ranges. We will, nonetheless keep $c_1-c_2+c_5=0$ to avoid the leading powers violating the asymptotic behaviour \cite{Dumm:2012vb}. In this way, we 
will assume variations of $\pm 20\%$ for the non-vanishing combinations of couplings in eq.(\ref{Couplings}): $c_6-c_5$, $d_1+8d_2-d_3$, $d_3$, $G_V$, $F_V$, $F_A$, $\lambda'$ 
and $\lambda_0$ and we will set $|c_1+c_2+8c_3-c_5|\leq0.01$ and $|\lambda''|\leq0.04$ so that they are smaller than analogous non-vanishing couplings according to 
eq.(\ref{Couplings}).

\section{Phenomenological analysis}\label{Pheno}

Using the results of previous sections, we have evaluated the branching fractions and the invariant mass spectrum of the $\ell^+\ell^-$ pair for the decays 
$\tau^-\to\pi^-\nu_\tau\ell^+\ell^-$ ($\ell=e,\,\mu$). In order to assess the contributions of structure-dependent ($SD$) and inner-bremsstrahlung ($IB$) contributions we 
have evaluated separately the moduli squared and interferences in both observables as discussed in Section 2. 
The form factors that describe $SD$ contributions were given in eqs.(\ref{F_V}) to (\ref{loopfun_2pi}) and the coupling constants involved were fixed using short-distance $QCD$ 
constraints in eq.(\ref{Couplings}). The branching ratios that are predicted using these form factors are shown in the second and third columns of Table \ref{Tab:1}; the 
corresponding allowed ranges that are obtained by letting the couplings to vary within $20\%$ of their central values, as described in the previous section, are shown in the 
fourth and fifth columns of Table \ref{Tab:1}. The couplings which were predicted to vanish ($c_1+c_2+8c_3-c_5$ and $\lambda''$) have a marginal influence on the error estimates. 
Also, the impact of the variations on $\lambda_0$, $\lambda'$ and on $d_1+8d_2-d_3$ are rather mild and the error ranges are basically determined by the uncertainties on the 
remaining couplings: $F_V$, $F_A$, $G_V$, $c_5-c_6$ and $d_3$.
\begin{table*}[h!]
 \begin{center}
\begin{tabular}{|c||c|c||c|c|}
\hline
 & $\ell=e$ & $\ell=\mu$& $\ell=e$ & $\ell=\mu$\\
\hline
IB& $1.461\cdot10^{-5}$ & $1.600\cdot10^{-7}$ & $\pm 0.006\cdot10^{-5}$& $\pm 0.007\cdot10^{-7}$\\
IB-V& $-2\cdot10^{-8}$ & $1.4\cdot10^{-8}$ & $\left[-1\cdot10^{-7},1\cdot10^{-7}\right]$ & $\left[-4\cdot10^{-9},4\cdot10^{-8}\right]$\\
IB-A& $-9\cdot10^{-7}$ & $1.01\cdot10^{-7}$ & $\left[-3\cdot10^{-6},2\cdot10^{-6}\right]$ & $\left[-2\cdot10^{-7},6\cdot10^{-7}\right]$\\
VV & $1.16\cdot10^{-6}$ & $6.30\cdot10^{-7}$ & $\left[4\cdot10^{-7},4\cdot10^{-6}\right]$ & $\left[1\cdot10^{-7},3\cdot10^{-6}\right]$\\
AA& $2.20\cdot10^{-6}$ & $1.033\cdot10^{-6}$ & $\left[1\cdot10^{-6},9\cdot10^{-6}\right]$ & $\left[2\cdot10^{-7},6\cdot10^{-6}\right]$\\
V-A& $2\cdot10^{-10}$ & $-5\cdot10^{-11}$ & $\sim10^{-10}$  & $\sim10^{-10}$\\
\hline
TOTAL& $1.710\cdot10^{-5}$& $1.938\cdot10^{-6}$ & $\left(1.7^{+1.1}_{-0.3}\right)\cdot 10^{-5}$& $\left[3\cdot10^{-7},1\cdot10^{-5}\right]$\\
\hline
\end{tabular}
\caption{\small{The central values of the different contributions to the branching ratio of the $\tau^-\to\pi^-\nu_\tau\ell^+\ell^-$ decays ($\ell=e,\,\mu$) are displayed 
on the left-hand side of the table. The error bands of these branching fractions are given in the right-hand side of the table. The error bar of the IB contribution stems 
from the uncertainties on the $F_{\pi}$ decay constant and $\tau$ lepton lifetime \cite{Beringer:1900zz}.}} \label{Tab:1}
\end{center}
\end{table*}

\begin{figure}[h!]
\centering
\vspace{1.3cm}
\includegraphics[scale=0.55,angle=-90]{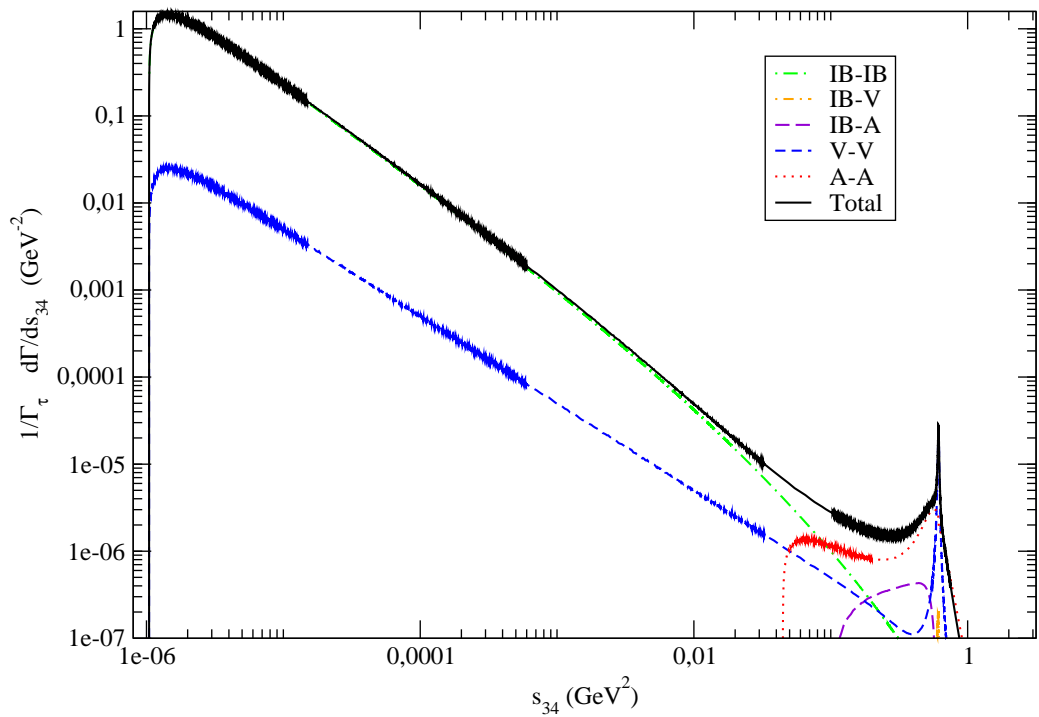}
\caption{\small{The different contributions to the normalized $e^+e^-$ invariant mass distribution defined in Eq. (\ref{spectrum}) are plotted. A double logarithmic scale was 
needed.} \label{Fig:4}}
\end{figure}

\begin{figure}[h!]
\includegraphics[scale=0.55,angle=-90]{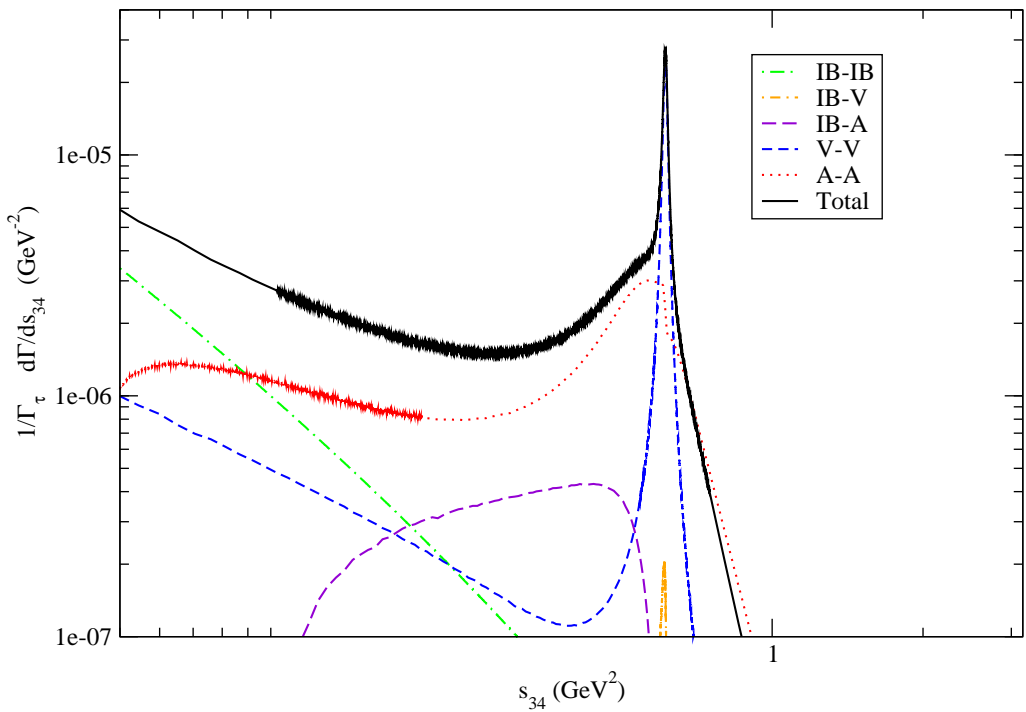}
\caption{\small{The different contributions to the normalized $e^+e^-$ invariant mass distribution defined in Eq. (\ref{spectrum}) are plotted in a magnification for 
$s_{34}\gtrsim 0.1$ GeV$^2$ intended to better appreciate the $SD$ contributions. A double logarithmic scale was needed.} \label{Fig:5}}
\end{figure}

The normalized invariant-mass distribution of the lepton pair,
\begin{equation}\label{spectrum}
\frac{1}{\Gamma_\tau}\cdot \frac{d\Gamma\left(\tau^-\to\pi^-\nu_\tau e^+e^-\right)}{ds_{34}}\, ,
\end{equation}
 is shown in Fig. \ref{Fig:4}. As it can be observed, the $IB$ contribution dominates the spectrum for values of $s_{34}\lesssim 0.1$ GeV$^2$. For larger 
values (which can be better appreciated in Fig. \ref{Fig:5}) the $SD$ part overcomes the former and the $AA$ contribution dominates in the rest of the spectrum apart from 
the $\rho(770)$ peak region where the $VV$ part overtakes it. The interference terms $IB-V$ and $IB-A$ are negative for most of the spectrum and do not appear in the 
figure.

\begin{figure}[!h]
\begin{center}
\vspace*{1.2cm}
\includegraphics[scale=0.55,angle=-90]{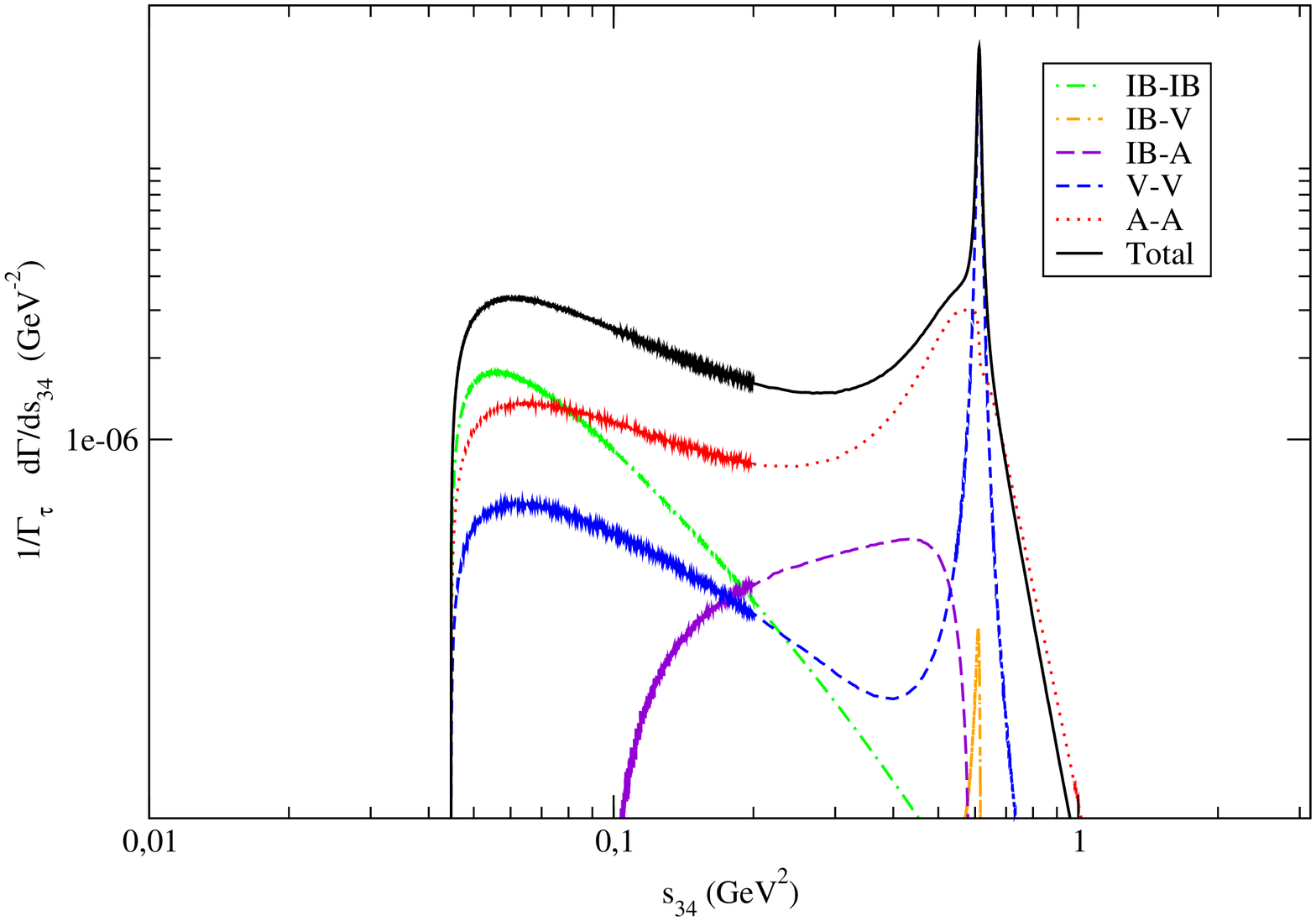}
\caption[]{\small{The different contributions to the normalized $\mu^+\mu^-$ invariant mass distribution are plotted. A double 
logarithmic scale allows to display the different contributions more clearly.}} \label{Fig:6}
\end{center}
\end{figure}

The normalized $\mu^+\mu^-$ invariant mass distribution (similar definition as in Eq. (\ref{spectrum})) is shown in Fig. \ref{Fig:6}. In this case the $IB$ and 
$SD$ contributions (essentially $AA$ apart from the $\rho(770)$ peak region) are comparable for $s_{34}\lesssim 0.1$ GeV$^2$. For higher values of the squared photon invariant 
mass the main contribution comes from the $AA$ part and the $VV$ contribution shows up through the peak at the $\rho(770)$ mass.

In Figs.~\ref{Fig:4}-\ref{Fig:6} vertical fluctuations can be appreciated in certain energy regions of the normalized invariant-mass distributions. In order to compute these 
distributions in the $s_{34}$ variable, we have integrated numerically the decay probability over the remaining four independent kinematical variables by using a fortran code 
based on the VEGAS routine. The observed fluctuations  arise from the Monte Carlo evaluation over the four-body phase space integration. The branching fractions shown in Table 
\ref{Tab:1} were obtained by integrating numerically these invariant-mass distributions and checked from a direct integration over the five independent kinematical variables.

We have found that the $SD$ contribution is sizable ($15\%$) in the case of $\ell=e$ decays and dominant ($92\%$) for $\ell=\mu$. Accordingly, it will be easy to 
pin it down from the experimental data if enough statistics is accumulated: in $\ell=e$ decays by confirming that the differential decay width ceases to decrease as expected 
from $IB$ around $s_{34}\sim 0.1$ GeV$^2$ and starts increasing up to the $\rho(770)$ peak region; in the $\ell=\mu$ case first because if falls down slower than expected 
from a $QED$ contribution \footnote{The $(1/\Gamma)\, d\Gamma/ds_{34}$ distribution and the $IB$ contribution to it can be well approximated by $a+b$ Log$(s_{34})$ in the range 
$\left[0.11,0.19\right]$ GeV$^2$. We find $b^{TOT}=-1.314(3)\cdot10^{-6}$ and $b^{IB}=-8.87(3)\cdot10^{-7}$, quantifying the effect of $SD$ contributions in this region. We 
quote for completeness our results $a^{TOT}=-5.63(6)\cdot10^{-7}$ and $a^{IB}=-1.221(5)\cdot10^{-6}$.} and, from $s_{34}\sim 0.3$ GeV$^2$ on, because it starts to rise up 
to the $\rho(770)$ peak region. In case a fine binning is achieved in this zone it will be possible to confirm the expected $VV$ contribution in either decay mode as well.

The fact that, in both decays, the contribution to the decay width of the $s_{34}>1$ GeV$^2$ region is negligible justifies our assumption of including only the lightest 
multiplet of vector and axial-vector resonances. This result is not trivial in the axial-vector case and in the vector case it is not modified even if the 
$\rho(1450)$ exchange is included phenomenologically \cite{Dumm:2009va}.

We have also assessed the relevance of the axial-vector $B$ form factor, introduced in eq.\ref{A matrix element} (see also eq.\ref{B}). We find it important, 
as the $(AA)+(IB-A)$ contributions drop to $33\%$ and $25\%$ of the values shown in Table \ref{Tab:1} if this form factor is neglected. This, in turn, results in a decrease 
of the branching ratio of $5\%$ for $\ell=e$ and $44\%$ for $\ell=\mu$. Therefore, it is essential to include this contribution in the muon decay channel. This explains why 
the $AA$ normalized invariant mass distribution was peaked in the $\rho(770)$ mass region for either channel, since the $B$ form factor is proportional to the isovector 
component of the electromagnetic di-pion form factor.

A future study of the data corresponding to the $SD$-dominated part of the spectrum will also allow to test the hadronization proposed in Ref.~\cite{Guo:2010dv} for the 
$\tau^-\to\pi^-\gamma\nu_\tau$ decays. In particular, in that reference it was found that
\begin{equation}
 \Gamma\left(\tau^-\to\pi^-(\gamma)\nu_\tau\right)=\Gamma\left(\tau^-\to\pi^-\nu_\tau\right)(1+\delta_\gamma)\,,
\end{equation}
with $\delta_\gamma\sim1.460\cdot10^{-2}$ for a photon energy threshold of $50$ MeV. The $SD$ part, whose contribution was found to be $\delta_\gamma\sim0.138\cdot10^{-2}$, 
could be tested through the $\tau^-\to\pi^-\nu_\tau\ell^+\ell^-$ ($\ell=e,\,\mu$) decays considered in this paper. This knowledge can also be extended to the computation of 
the radiative corrections to the ratio $R_{\tau/\pi}:=\Gamma\left(\tau^-\to\pi^-\nu_\tau\right)/\Gamma\left(\pi^-\to\mu^-\bar{\nu}_\mu\right)$ \cite{radtaudec}, relevant 
for lepton universality tests \cite{Pich:2013kg}.

Finally, the study of radiative tau decays is also important for a faithful modelling of backgrounds in lepton flavour violation searches, as it was noted for the 
$\tau^-\to\pi^-\gamma\nu_\tau$ decays in the case where the pion is missidentified as a muon and resembles the $\tau^-\to\mu^-\gamma$ \cite{Guo:2010ny} signal. The standard 
simulation of the radiative decay is performed with PHOTOS \cite{PHOTOS}, which only includes the scalar $QED$ contribution neglecting the $SD$ parts. Analogously, the 
$\tau^-\to\pi^-\ell^+\ell^-\nu_\tau$ ($\ell=e,\,\mu$) decays under consideration might also mimick the $\tau^-\to\mu^-\ell^+\ell^-$ processes. Although it seems that 
the inclusion of $QCD$ contributions for the $\ell=\mu$ case will be important (as the $SD$ part gives the bulk of the branching ratio), a devoted study is needed to confirm 
this, because the involved processes are three- and four-body decays, which complicates things with respect to the study in Ref.~\cite{Guo:2010ny}, where the kinematics of 
$\tau^-\to\mu^-\gamma$ is completely fixed selecting the photons with almost maximal energy in $\tau^-\to\pi^-\gamma\nu_\tau$ decays as the relevant background.

\section{Conclusions}\label{Concl}
We have studied for the first time the $\tau^-\to\pi^-\nu_\tau\ell^+\ell^-$ ($\ell=e,\,\mu$) decays. We have evaluated the model-independent contributions by using $QED$ 
and have obtained the structure-dependent part ($W^* \to \pi^-\gamma^*$ vertex) using $R\chi T$. This approach ensures the low-energy limit of $\chi PT$ and includes the 
lightest resonances as active degrees of freedom worked out within the convenient antisymmetric tensor formalism. We have been able to predict all the couplings involved in the 
relevant Lagrangian term using short-distance QCD constraints (in the $N_C\to\infty$ limit and restricting the spectrum to the lowest-lying spin-one resonances) on the 
related Green functions and form factors and considered the error stemming from this procedure in a conservative way. 

Within this framework we predict 
$BR\left(\tau^-\to\pi^-\nu_\tau e^+e^-\right)=\left(1.7^{+1.1}_{-0.3}\right)\cdot 10^{-5}$ and \break $BR\left(\tau^-\to\pi^-\nu_\tau \mu^+\mu^-\right)\in \left[3\cdot10^{-7},1\cdot10^{-5}\right]$. 
We find that while the $\ell=e$ decays should be within discovery reach at the future super-flavour facilities, this will only be possible for the $\ell=\mu$ decays if they 
happen to be close to the upper limit of the range we have given.
The studied hadronic currents are ready for installation in the $R\chi T$ based version \cite{TAUOLA2, Shekhovtsova:2012ra} of TAUOLA, the standard Monte Carlo generator for 
tau lepton decays.
\section*{Acknowledgements} This work has been partially supported by the Spanish grant FPA2011-25948 and Conacyt (M\'exico). P.R. acknowledges the hospitality of 
Departamento de F\'{\i}sica at CINVESTAV, where part of this work was done.

\section*{Appendix} \label{LongExpressions}
We collect in this appendix the results of summing over polarizations and averaging over that of the tau the different contributions to the squared matrix element. We refrain 
from writing the lengthy outcome of the contraction of the indices which was used in our programs.
\begin{eqnarray}\label{averaged IB}
& & \overline{\Big|\mathcal{M}_{IB}\Big|^2} = 16 G_F^2 |V_{ud}|^2 \frac{e^4}{k^4}F_\pi^2 M_\tau^2 \ell_{\mu\nu} \left[\frac{-\tau^{\mu\nu} k^2}
{\left(k^2-2 k\cdot p_\tau\right)^2}+\frac{4 p^{\mu} q^{\nu} k\cdot p_\tau}{\left(k^2+2 k\cdot p\right) \left(k^2-2 k\cdot p_\tau\right)}\right.\nonumber\\
& & \left. +\frac{4 p_\tau^{\mu} q^\nu k\cdot p_\tau}{\left(k^2-2 k\cdot p_\tau\right)^2}-\frac{2 g^{\mu \nu} k\cdot p_\tau k\cdot q}{\left(k^2-2 k\cdot p_\tau\right)^2}-\frac{4 p^{\mu} p_\tau^{\nu} k\cdot q}{\left(k^2+2 k\cdot p\right) \left(k^2-2 k\cdot p_\tau\right)}\right.\nonumber\\
& & \left. -\frac{4 p_\tau^{\mu} p_\tau^{\nu} k\cdot q}{\left(k^2-2 k\cdot p_\tau\right)^2}+\frac{8 p^{\mu} p_\tau^{\nu} p_\tau\cdot q}{\left(k^2+2 k\cdot p\right) \left(k^2-2 k\cdot p_\tau\right)}\right.\nonumber\\
& & \left. +\frac{4 p^{\mu} p^{\nu} p_\tau\cdot q}{\left(k^2+2 k\cdot p\right)^2}\,+\frac{4 p_\tau^{\mu} p_\tau^{\nu} p_\tau\cdot q}{\left(k^2-2 k\cdot p_\tau\right)^2}\right]\,,\nonumber\\
\end{eqnarray}
\begin{eqnarray}\label{averaged IBV}
& & \overline{2 \Re e\left[\mathcal{M}_{IB}\mathcal{M}_V^*\right]} = - 32  G_F^2|V_{ud}|^2 \frac{e^4}{k^4}F_\pi M_\tau^2 \Im m \left\lbrace F_V^*(p\cdot k,k^2)\ell^\mu_{\nu^\prime} \epsilon^{\mu^\prime \nu^\prime \rho^\prime \sigma^\prime}k_{\rho^\prime}p_{\sigma^\prime} \mathcal{V}_{\mu\mu^\prime}\right\rbrace\,,\nonumber
\end{eqnarray}
\begin{eqnarray}\label{averaged IBA}
& &  \overline{2 \Re e\left[\mathcal{M}_{IB}\mathcal{M}_A^*\right]} = -64 G_F^2 |V_{ud}|^2 \frac{e^4}{k^4}F_\pi M_\tau^2 \ell_\mu^{\nu^\prime} \Re e\left[ \mathcal{A}^*_{\mu^\prime\nu^\prime} \mathcal{V}^{\mu\mu^\prime}\right]\,,\nonumber
\end{eqnarray}
\begin{eqnarray}\label{averaged V}
& & \overline{ \Big|\mathcal{M}_{V}\Big|^2} = 16 G_F^2 |V_{ud}|^2 \frac{e^4}{k^4} \Big|F_V(p\cdot k,k^2)\Big|^2\epsilon_{\mu^\prime\nu^\prime\rho^\prime\sigma^\prime}\epsilon_{\mu\nu\rho\sigma}
k^\rho p^\sigma k^{\rho\prime} p^{\sigma\prime} \ell^{\nu{\nu^\prime}}\tau^{\mu{\mu^\prime}}\,,\nonumber
\end{eqnarray}
\begin{eqnarray}\label{averaged A}
& & \overline{\Big|\mathcal{M}_{A}\Big|^2} = 64 G_F^2 |V_{ud}|^2 \frac{e^4}{k^4}\ell_{\nu{\nu^\prime}}\tau_{\mu{\mu^\prime}}\mathcal{A}^{\mu\nu}{\mathcal{A}^{\mu^\prime\nu^\prime}}^*\,,\nonumber
\end{eqnarray}
\begin{eqnarray}\label{averaged VA}
& &  \overline{2 \Re e\left[\mathcal{M}_{V}\mathcal{M}_A^*\right]} = -64  G_F^2 |V_{ud}|^2 \frac{e^4}{k^4}\Im m\left[ F_V(p\cdot k,k^2)\epsilon_{\mu\nu\rho\sigma}k^\rho p^\sigma \ell_{\nu^\prime}^\mu \tau^{\mu\mu^\prime}{\mathcal{A}_{\mu^\prime}^{\nu^\prime}}^*\right]\nonumber\,,
\end{eqnarray} 
where we have defined
\begin{eqnarray}
 \ell^{\mu\nu} & = & p_-^\mu p_+^\nu+p_-^\nu p_+^\mu-g^{\mu\nu}(m_\ell^2+p_-\cdot p_+)\, \nonumber \\  \tau^{\mu\nu}& =& p_\tau^\mu q^\nu+p_\tau^\nu q^\mu-g^{\mu\nu}p_\tau \cdot q\,,\\
 \mathcal{A}^{\mu\nu}& = & F_A(p\cdot k,k^2)\left[(k^2+p\cdot k)g^{\mu\nu}-k^\mu p^\nu\right]+B(k^2) k^2 \left[g^{\mu\nu}-\frac{(p+k)^\mu p^\nu}{k^2+2p\cdot k}\right]\,,\nonumber\\
 \mathcal{V}_{\mu\nu} & = & \frac{2p_\mu q_{\nu}}{2k\cdot p+k^2}+\frac{-g_{\mu\nu}k\cdot q+2q_{\nu}p_{\tau\,\mu} -i\epsilon_{\mu\nu\rho\sigma}k^\rho q^\sigma+k_{\nu}q_\mu}{k^2-2k\cdot p_\tau}\, ,\nonumber
\end{eqnarray}\label{definitions}
and used the conservation of the electromagnetic currents implying $k_\mu\ell^{\mu\nu}=0=\ell^{\mu\nu}k_\nu$.

\end{document}